# Non-Empirically Tuned Range-Separated DFT Accurately Predicts Both Fundamental and Excitation Gaps in DNA and RNA Nucleobases

*Michael E. Foster and Bryan M. Wong\**

Materials Chemistry Department, Sandia National Laboratories, Livermore, California 94551

*Corresponding author. E-mail: bmwong@sandia.gov. Homepage: http://alum.mit.edu/www/usagi

Using a non-empirically tuned range-separated DFT approach, we study both the quasiparticle properties (HOMO-LUMO fundamental gaps) and excitation energies of DNA and RNA nucleobases (adenine, thymine, cytosine, guanine, and uracil). Our calculations demonstrate that a physically-motivated, first-principles tuned DFT approach accurately reproduces results from both experimental benchmarks and more computationally intensive techniques such as many-body GW theory. Furthermore, in the same set of nucleobases, we show that the non-empirical range-separated procedure also leads to significantly improved results for excitation energies compared to conventional DFT methods. The present results emphasize the importance of a non-empirically tuned range-separation approach for accurately predicting both fundamental and excitation gaps in DNA and RNA nucleobases.

**Introduction**

DNA and RNA nucleobases contain the genetic information of all living cells and play a vital role in the development and functioning of all known living organisms. Because of their importance in maintaining the genome integrity during cell replication, increasing attention continues to be devoted to understanding the specific mechanisms (particularly ionization and electron-impact radiation damage) that can dramatically alter their electronic and structural states. An important first step in understanding the reactivity and damage mechanisms in DNA and RNA is the ability to both *efficiently* and *accurately* predict electronic properties such as ionization energies and electron affinities. While numerous theoretical studies using high-level wavefunction-based techniques have been used to accurately predict the electronic properties of nucleobases, (see Ref. 1, and references within), their immense computational costs prevent their routine use for complex biological environments or for large geometries (i.e., a fully-periodic DNA/RNA strand is currently not possible with wavefunction-based methods). Because of its favorable balance between accuracy and efficiency, density functional theory (DFT) has become the most widely used quantum mechanical method for obtaining electronic structure information of molecules and solids. However, selecting the best exchange-correlation functional continues to be a difficult task due to the large number of functionals available.

Herein, we assess the accuracy of a non-empirically tuned range-separated (long-range corrected) density functional theory (LC-DFT) method[2-4] for predicting both the quasiparticle properties (HOMO-LUMO fundamental gaps) and the excitation energies in DNA and RNA nucleobases. The fundamental gap is rigorously defined as the difference in energy between the first ionization potential (IP) and the first electron affinity (EA) whereas the excitation/optical gap is the difference in energy between the lowest dipole-allowed excited state and the ground state. In general, the true optical excitation energy is smaller than the fundamental gap due to excitonic effects that arise from the Coulombic attraction between the excited electron and hole within the molecule (see Fig.7 in Ref. 5 for a clear pictorial example of the fundamental gap and exciton binding energy). As discussed extensively by Izmaylov and Scuseria,[6] semilocal functionals are incapable of accurately predicting excitonic effects whereas exchange-correlation kernels that include a portion of nonlocal long-range exact exchange (such as the LC-DFT kernels used in this work) give rise to a non-zero exciton binding energy, resulting in an accurate prediction of the optical gap. In this study, we investigate the quasiparticle properties for guanine, adenine, cytosine, thymine, and uracil (molecular structures shown in Fig. 1), and we evaluate the performance of the non-empirically tuned LC-DFT approach against both wavefunction-based and conventional DFT methods. Finally, using the same non-empirical tuning procedure, we assess the accuracy of this approach and discuss the implications for predicting the excited-state energies and properties in these systems.

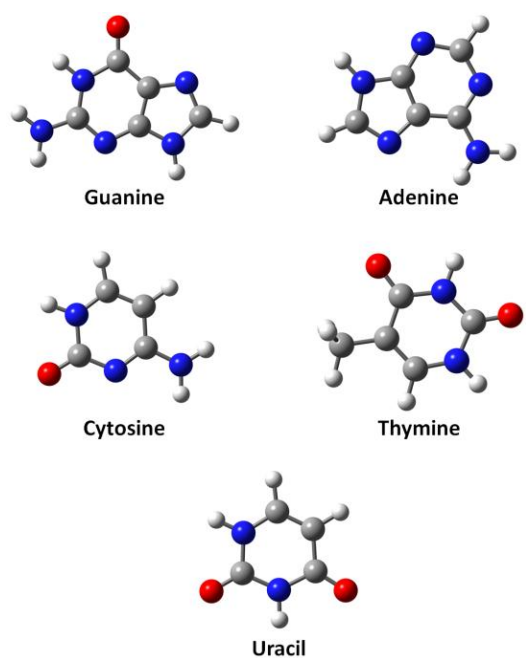

**Figure 1.** Molecular structures of the DNA/RNA nucleobases

**Theory and Methodology**

Over the last few years, the use of range-separated functionals[2-4,6-16] for both DFT and time-dependent DFT (TDDFT) applications has significantly grown in popularity. In particular, we and other researchers have shown that these functionals show a dramatic improvement for strong charge-transfer systems[3,9,10] and also surprisingly show improved accuracy even for relatively simple valence excitations.[11,12,14] In contrast to conventional hybrid functionals which incorporate a constant fraction of Hartree-Fock exchange, the range-separated formalism mixes exchange densities nonuniformly by partitioning the electron repulsion operator into short-range (1st term) and long-range (2nd term) contributions as:

$$\frac{1}{r_{12}} = \frac{1-\text{erf}(\mu \cdot r_{12})}{r_{12}} + \frac{\text{erf}(\mu \cdot r_{12})}{r_{12}}. \quad (1)$$

The "erf" term denotes the standard error function, $r_{12}$ is the interelectronic distance between electrons 1 and 2, and $\mu$ is the range-separation parameter in units of Bohr$^{-1}$. For a pure density functional (i.e., BLYP or PBE) which does not already include a fraction of nonlocal Hartree-Fock exchange, the exchange-correlation energy according to the LC formalism is

$$E_{xc}(\mu) = E_{c,\text{DFT}}(\mu) + E_{x,\text{DFT}}^{\text{SR}}(\mu) + E_{x,\text{HF}}^{\text{LR}}(\mu) \quad (2)$$

where $E_{c,\text{DFT}}$ is the DFT correlation functional, $E_{x,\text{DFT}}^{\text{SR}}$ is the short-range DFT exchange functional, and $E_{x,\text{HF}}^{\text{LR}}$ is the Hartree-Fock contribution to exchange computed with the long-range part of the Coulomb operator. The modified (nonlocal) $E_{x,\text{HF}}^{\text{LR}}$ term can be analytically evaluated with Gaussian basis functions, and the short-range $E_{x,\text{DFT}}^{\text{SR}}$ contribution is computed with a modified exchange kernel specific for each generalized gradient approximation.[7] Recently, Baer, Kronik,[2-4] and others[15,16] have demonstrated that the range-separation parameter in these exchange-correlation kernels is highly system dependent but can be non-empirically tuned for a given system. This tuning process ensures that the negative of the HOMO energy is equal to the ionization potential (IP) of the $N$ electron system, which is a fundamental condition within the Kohn-Sham DFT formalism (i.e., Janak's theorem[17]). This condition would naturally be satisfied if the exact exchange-correlation functional were known; as a result, tuning $\mu$ in a self-consistent manner to satisfy this fundamental constraint is both intuitive and theoretically justified. More rigorously, the IP of a given system can be determined from the difference between the ground-state energy of the *N* electron and the *N*-1 electron systems. This energy difference corresponds to the energy required to remove an electron from the system and, according to Janak's theorem should be equal to the negative of the HOMO energy. As mentioned, this condition can be satisfied (or approximately satisfied) with an optimal range-separation parameter that can be obtained by minimizing the following function:[3]

$$J^2(\mu) = \left[\varepsilon_{HOMO}^{\mu}(N) + \mathrm{IP}^{\mu}(N)\right]^2 + \left[\varepsilon_{HOMO}^{\mu}(N+1) + \mathrm{IP}^{\mu}(N+1)\right]^2. \qquad (3)$$

The second term in this function takes into consideration the *N*+1 system to indirectly *tune* the LUMO of the *N* electron system. The LUMO cannot be directly incorporated in Eq. 3 as there is no formal equivalent of Janak's theorem for the LUMO; that is, Janak's theorem does not explicitly relate the electron affinity to the negative of the LUMO. As we will see, the tuning of $\mu$, besides being theoretically rigorous, significantly improves the ability to predict HOMO and LUMO levels, fundamental gaps, and even excitation energies.

In order to maintain a consistent comparison across previously published benchmark calculations, identical molecular geometries obtained from Ref. 1 were used for this work. These reference geometries are available in the Supporting Information. Optimal $\mu$ values were determined for guanine, adenine, cytosine, thymine, and uracil with the long-range corrected BLYP functional[7] (LC-BLYP) using a polarized triple zeta basis set (cc-pVTZ). We also investigated the effect of including diffuse functions using a larger aug-cc-pVTZ basis, but found that the use of larger or more diffuse basis sets did not significantly change our overall findings (see Results and Discussion section for further details). The optimal range-separation values were determined by varying $\mu$ from 0.05 → 0.9 in increments of 0.05 (increments of 0.01 were used from 0.2 → 0.4). In order to determine $J^2$, single-point calculations were carried out for the *N* and *N* ± 1 states for each $\mu$ value. Figure 2 graphically illustrates $J^2$ as a function of $\mu$ for the different nucleobases. The minimum (optimal $\mu$) of each curve was obtained by spline interpolation, and these optimal values are reported in Table 1. These optimal range-separation parameters were used for all subsequent LC-BLYP calculations. All calculations were carried out with the Gaussian 09 package[18] using default SCF convergence criteria (density matrix converged to at least $10^{-8}$) and the default DFT integration grid (75 radial and 302 angular quadrature points).

| Nucleobase | Optimal $\mu$ (Bohr$^{-1}$) |
|---|---|
| Guanine | 0.2738 |
| Adenine | 0.2853 |
| Cytosine | 0.2948 |
| Thymine | 0.2850 |
| Uracil | 0.3060 |

**Table 1.** LC-BLYP/cc-pVTZ optimal $\mu$ values for the different DNA/RNA nucleobases

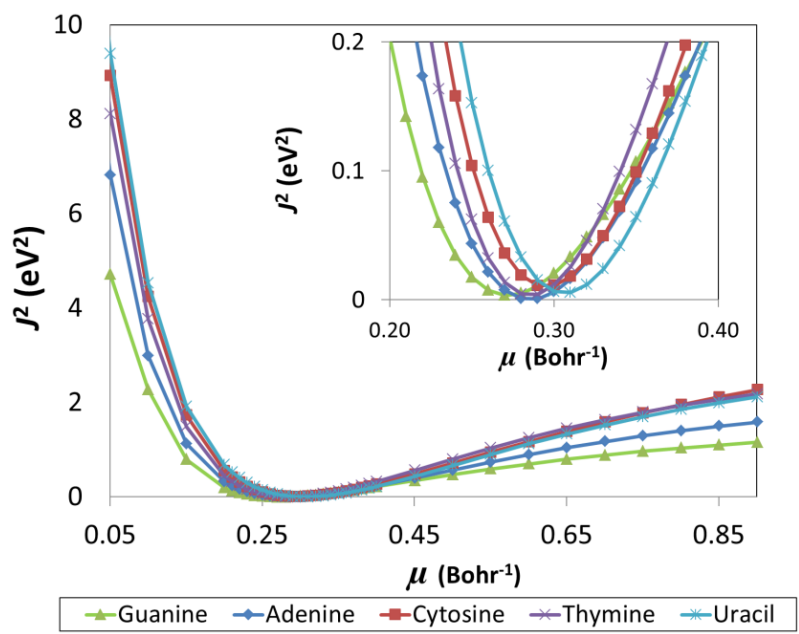

**Figure 2.** $J^2$ (Eq. 3) as a function of $\mu$ for the different DNA/RNA nucleobases as determined using the LC-BLYP functional and cc-pVTZ basis. The inset shows a magnified view of $J^2$ in the $0.2 < \mu < 0.4$ range.

**Results and Discussion**

The ability of the non-empirically tuned LC-BLYP method to predict fundamental gaps and molecular orbital energies is investigated by comparing results to recently reported GW, CASPT2, and experimental values.[1] The GW method[19,20] is based on a Green's function formalism which self-consistently corrects the Kohn-Sham eigenvalues to give significant improvements in the orbital energies. The CASPT2 (complete active-space second-order perturbation theory) method[21] is a multiconfigurational perturbation technique which gives an estimate to the full configuration interaction (CI) energy. The theoretically predicted HOMO-1, HOMO, and LUMO (eV) levels and the computed fundamental gaps (LUMO–HOMO) for the 5 nucleobases considered are reported in Table 2. It should be noted that the "HOMO" and "LUMO" quasiparticle energies are not well-defined quantities in wavefunction-based CASPT2 calculations, and the reported HOMO/LUMO energies in Table 2 are, in fact, vertical ionization energies and electron affinities, respectively. In addition to the mentioned methods, results using the popular B3LYP[22] and the standard LDA functionals are also reported.

Using the CASPT2 method as our benchmark, we find that the non-empirically tuned LC-BLYP method produces the overall best results, even outperforming the GW method. The GW method predicts the HOMO energy levels slightly better; however, the non-empirically tuned LC-BLYP method is more accurate for the HOMO-1 energy levels (0.42 and 0.02 eV MAE, respectively) and the fundamental gaps (0.12 and 0.06 eV MAE, respectively). The LC-BLYP method even slightly outperforms for predicting LUMO energy levels. As mentioned in the Theory and Methodology section, we also investigated the effect of including diffuse functions, and found that our LC-BLYP/aug-cc-pVTZ results actually show a better agreement with experiment while deviating slightly more from the GW and CASPT2 benchmarks (see Tables S1 and S2 in the Supporting Information). We attribute the deviations between the LC-BLYP/aug-cc-pVTZ and GW/CASPT2 calculations to the non-diffuse basis functions used in the computationally demanding wavefunction-based methods. For this reason, we only show the cc-pVTZ calculations in Table 2 since this comparison allows a fair and consistent evaluation since all of these theoretical methods use similarly sized basis sets. A quick analysis of Table 2 shows that both the LDA and B3LYP methods perform extremely poorly for predicting orbital energy levels and fundamental gaps compared to the GW and LC-BLYP methods. Strikingly, in all cases, both the LDA and B3LYP calculations are *qualitatively incorrect*, predicting the *wrong sign of the LUMO*. The mean absolute errors (MAE) for the molecular orbitals and fundamental gaps are reported in Table 2. The LC-BLYP method produces results very comparable to the more rigorous GW method, and the MAE between the two methods is only 0.06 eV for predicting the fundamental gaps. Furthermore, the non-empirically tuned LC-BLYP functional significantly outperforms LDA and B3LYP, and qualitative improvements are even achieved.

It is important to mention at this point two additional aspects that make the non-empirical tuning procedure more efficient than the computationally demanding wavefunction-based methods. First, although we calculated several single-point energies to determine the optimal range-separation parameter (i.e., 37 different $\mu$ values were used for each nucleobase), we found that substantially fewer calculations are actually required, and a coarse grid of $\mu$ values (in increments of ~ 0.05 or even larger) fit to a smooth spline gives nearly identical results. The large number of calculations used to generate Fig. 2 was carried out for completeness. Second, it is important to note that even if one chooses to use a fine grid of $\mu$ values, each of the 37 different calculations is completely independent of each other and can actually be calculated separately on different CPUs. In contrast, the prohibitive computational scaling of wavefunction-based methods (the EOM-CCSD formalism used later in this work scales as $N^6$) prevents their routine use even if several processors are used in parallel. For example, a single LC-BLYP/cc-pVTZ calculation for guanine takes less than an hour on 8 × 2.93 Ghz Intel Nehalem CPUs whereas a calculation at the EOM-CCSD/cc-pVTZ level of theory takes over 5 days using the same computational resources. This combination of efficiency with the accuracy demonstrated previously demonstrates a clear benefit of the non-empirically tuned range-separated formalism.

|  | LDA-KS[†] | B3LYP/cc-pvtz | LC-BLYP/cc-pvtz | GW[†] | CASPT2[†,‡] | Experiment[†] |
|---|---|---|---|---|---|---|
| Guanine | | | | | | |
| HOMO-1 | 6.34 | 7.11 | 9.29 | 9.82 | 9.56 | 9.9 |
| HOMO | 5.69 | 5.75 | 7.78 | 7.81 | 8.09 | 8.0 - 8.3 |
| LUMO | 1.8 | 0.30 | -1.69 | -1.58 | -1.14 | |
| Fundamental gap | 3.89 | 5.45 | 9.47 | 9.39 | 9.23 | |
| Adenine | | | | | | |
| HOMO-1 | 6.28 | 6.98 | 9.21 | 9.47 | 9.05 | 9.45 |
| HOMO | 6.02 | 6.11 | 8.21 | 8.22 | 8.37 | 8.3 - 8.5, 8.47 |
| LUMO | 2.22 | 0.81 | -1.13 | -1.14 | -0.91 | -0.56 to -0.45 |
| Fundamental gap | 3.80 | 5.30 | 9.33 | 9.36 | 9.28 | |
| Cytosine | | | | | | |
| HOMO-1 | 6.172 | 6.94 | 9.37 | 9.52 | 9.42 | 9.45, 9.55 |
| HOMO | 6.167 | 6.45 | 8.73 | 8.73 | 8.73 | 8.8 - 9.0, 8.89 |
| LUMO | 2.57 | 1.20 | -0.93 | -0.91 | -0.69 | -0.55 to -0.32 |
| Fundamental gap | 3.60 | 5.25 | 9.66 | 9.64 | 9.42 | |
| Thymine | | | | | | |
| HOMO-1 | 6.68 | 7.50 | 9.71 | 10.41 | 9.81 | 9.95 - 10.05, 10.14 |
| HOMO | 6.54 | 6.81 | 8.90 | 9.05 | 9.07 | 9.0 - 9.2, 9.19 |
| LUMO | 2.83 | 1.47 | -0.59 | -0.67 | -0.6 | -0.53 to -0.29 |
| Fundamental gap | 3.71 | 5.33 | 9.49 | 9.72 | 9.67 | |
| Uracil | | | | | | |
| HOMO-1 | 6.88 | 7.53 | 9.99 | 10.54 | 9.83 | 10.02 - 10.13 |
| HOMO | 6.72 | 7.15 | 9.45 | 9.47 | 9.42 | 9.4 - 9.6 |
| LUMO | 3.01 | 1.63 | -0.53 | -0.64 | -0.61 | -0.30 to -0.22 |
| Fundamental gap | 3.71 | 5.52 | 9.98 | 10.11 | 10.03 | |
| MAE HOMO-1 | 3.06 (3.48) | 2.32 (2.74) | 0.02 (0.44) | 0.42 | | |
| MAE HOMO | 2.51 (2.43) | 2.28 (2.20) | 0.12 (0.04) | 0.08 | | |
| MAE LUMO | 3.28 (3.47) | 1.87 (2.07) | 0.18 (0.02) | 0.20 | | |
| MAE Fundamental gap | 5.78 (5.90) | 4.16 (4.27) | 0.06 (0.06) | 0.12 | | |

**Table 2.** HOMO-1, HOMO, and LUMO energy levels (in eV) for the DNA/RNA nucleobases calculated at different levels of theory (note: the negative of the orbital energies are reported). The fundamental gap is determined by the difference between the HOMO and LUMO orbital energies. The MAE (mean absolute error) values are with respect to the CASPT2 values; the values in parentheses are with respect to the GW values. [†]Values were obtained from Ref. 1. [‡]The reported CASPT2 HOMO/LUMO energies are actually vertical ionization energies and electron affinities, respectively (see text).

The true test of any computational method is the ability to accurately reproduce experimental results. In Figure 3, the ionization energies determined by the various computational methods are graphically compared to the experimental ranges (numerical values are reported in Table 2). It is visually apparent that the CASPT2, GW, and LC-BLYP methods all produce results in good agreement with the experimental values; however, the LDA and the popular B3LYP methods show a dramatic underestimation of energies (errors > 2.0 eV). Again, this example demonstrates that traditional/popular density functionals can fail significantly for predicting ionization energies (-$\varepsilon_{HOMO}$); however, significant improvements can result from the use of non-empirically tuned LC-DFT functionals.

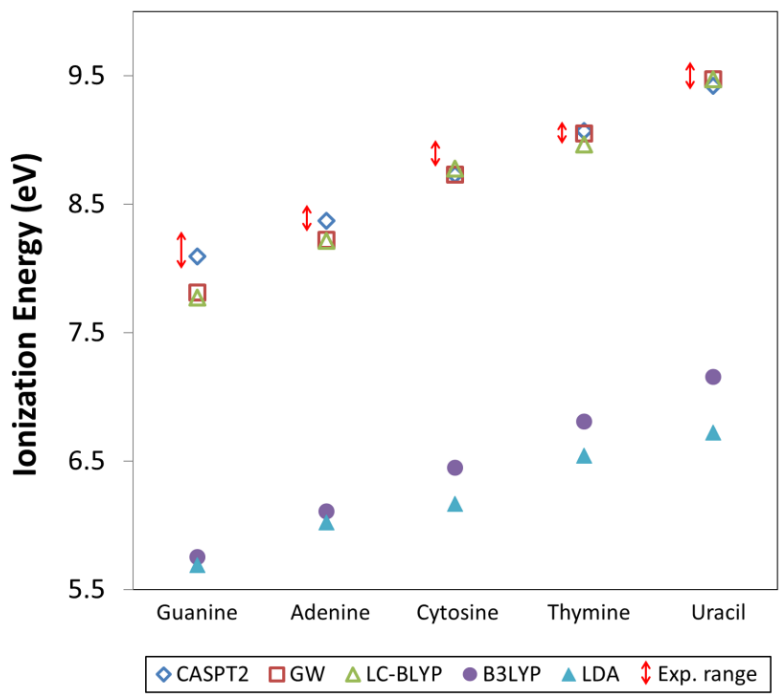

**Figure 3.** Ionization energies (eV) of the 5 nucleobases considered determined by the negative of the HOMO orbital at different levels of theory compared to the experimental range.

Finally, in addition to our detailed study of (ground-state) quasiparticle properties, we also analyzed the accuracy of non-empirically tuned range-separated methods for predicting optical excitation energies using time-dependent density functional theory (TDDFT). As benchmarks for assessing the quality of the various TDDFT methods, we calculated equation-of-motion couple-cluster[23] (EOM-CCSD) excitation energies with the same geometries and cc-pVTZ basis set used previously in our TDDFT calculations. It is important to mention that the lowest excited states in these nucleobases are largely dominated by $\pi \rightarrow \pi^*$ valence excitations and have been shown by the Krylov group to be well-described with the cc-pVTZ basis set.[24,25] For both the TDDFT and EOM-CCSD calculations, we also compute the oscillator strength since this provides another stringent benchmark test[26] for assessing excited-state properties. Table 3 compares both the lowest excitation energy and oscillator strengths between B3LYP, LC-BLYP, and EOM-CCSD for all 5 nucleobases. Overall, we find the non-empirically tuned LC-BLYP results are significantly in better agreement with the EOM-CCSD benchmark calculations in all cases. Specifically, Table 3 shows that the MAE is reduced *by more than half* for both the LC-BLYP excitation energies and oscillator strengths. In particular, we draw specific attention to the excitation energies of adenine, thymine, and uracil which are all significantly underestimated (by almost 0.5 eV) using B3LYP. In order to understand the possible cause of these dramatic errors, we calculated Tozer's lambda ($\Lambda$) diagnostic[27] at the B3LYP/cc-pVTZ level of theory for all 5 nucleobases. This test numerically quantifies the spatial overlap between the occupied and virtual orbitals involved in an excitation. By construction, the diagnostic metric $\Lambda$ is bounded between 0 and 1, with small values signifying a long-range excitation and large values indicating a localized, short-range transition. On the basis of their extensive benchmarks, if $\Lambda$ is less than 0.3, indicating little overlap and significant long-range charge transfer character, hybrid functionals are predicted to yield inaccurate results. As shown in the last column of Table 3, none of the $\Lambda$ values are less than 0.3; however, the results for adenine, thymine, and uracil are nearly borderline cases with smaller $\Lambda$ diagnostic values (~0.4). As a result, B3LYP exhibits larger errors for these particular systems when compared to our benchmark EOM-CCSD results. In contrast, it is remarkable to note that the non-empirically tuned LC-BLYP demonstrates significant accuracy for these excited states even though they have only been tuned to satisfy Kohn-Sham *ground-state* constraints (Eq. 3) *without relying on any a priori knowledge of excited-state properties*.

|   | B3LYP | | LC-BLYP | | EOM-CCSD | | |
|---|---|---|---|---|---|---|---|
| **Nucleobase** | $E_{abs}$/eV | Osc. Strength | $E_{abs}$/eV | Osc. Strength | $E_{abs}$/eV | Osc. Strength | Λ Diagnostic |
| Guanine | 4.89 | 0.122 | 5.00 | 0.134 | 5.19 | 0.166 | 0.75 |
| Adenine | 4.94 | 0.000 | 5.09 | 0.000 | 5.33 | 0.002 | 0.41 |
| Cytosine | 4.59 | 0.035 | 4.83 | 0.062 | 4.92 | 0.066 | 0.63 |
| Thymine | 4.66 | 0.000 | 4.91 | 0.001 | 5.15 | 0.000 | 0.41 |
| Uracil | 4.57 | 0.000 | 4.92 | 0.000 | 5.12 | 0.000 | 0.40 |
| MAE | 0.41 | 0.015 | 0.19 | 0.007 | | | |

**Table 3.** Excitation energies and oscillator strengths for the five nucleobases determined at different levels of theory. The lambda (Λ) diagnostic values were determined at the B3LYP/cc-pVTZ level of theory.

**Conclusion**

In conclusion, we have shown that non-empirically tuned range-separated DFT methods represent a significant improvement over traditional functionals for predicting both the quasiparticle properties (HOMO-LUMO fundamental gaps) and excitation energies in DNA and RNA nucleobases. We have demonstrated that the non-empirically tuned LC-BLYP method accurately reproduces experimental ionization potentials for the various nucleobases considered; in addition, our results demonstrate an excellent agreement with the more computationally intensive GW and CASPT2 methods. Furthermore, even though this non-empirical tuning procedure has been used to satisfy Kohn-Sham *ground-state* constraints, we have shown that this methodology also leads to significantly improved *excited-state* energies and properties in nucleobases, compared to conventional DFT methods. We believe that these types of functionals are at the forefront of density functional theory and will continue to grow in popularity because of their improved accuracy, computational efficiency, and theoretical rigorousness.

**Acknowledgement.** We thank Dr. Carina Faber and Dr. Xavier Blase for providing us with the Cartesian coordinates of the nucleobases studied here. B. M. W. acknowledges Dr. Pamela H. Li for her 30 years of stimulating ideas and inspiring discussions. Funding for this effort was provided by the Laboratory Directed Research and Development (LDRD) program at Sandia National Laboratories, a multiprogram laboratory operated by Sandia Corporation, a Lockheed Martin Company, for the United States Department of Energy under contract DEAC04-94AL85000.

**Supporting Information Available:** Benchmark calculations (Tables S1 and S2) with the aug-cc-pVTZ basis and Cartesian coordinates for all of the nucleobase structures. This material is available free of charge via the Internet at http://pubs.acs.org.